\documentclass[aps,nofootinbib,11pt,preprintnumbers]{revtex4-1}
\usepackage{amssymb,amsmath}
\usepackage{hyperref}
\usepackage{graphics}
\usepackage{graphicx}
\newcommand{\bea}{\begin{eqnarray}}
\newcommand{\eea}{\end{eqnarray}}
\newcommand{\beq}{\begin{equation}}
\newcommand{\eeq}{\end{equation}}

\newcommand{\eq}[1]{Eq.~(\ref{#1})}
\newcommand{\eqs}[1]{Eqs.~(\ref{#1})}
\newcommand{\ave}[1]{\langle {#1} \rangle}

\def\E{{\bf E}}
\def\F{{\bf F}}
\def\K{{\bf K}}
\def\0{{\bf 0}}

\def\nave{{\bar{n}}}

\def\epsave{{\bar{\epsilon}}}
\def\much{\mu_\mathit{c,hom}}
\def\muci{\mu_\mathit{c,inh}}
\def\ntyp{\ave{n}_\mathit{sol}}
\def\Rsperp{R_\mathit{sol,\perp}}
\def\Rsperprms{R_\mathit{sol,\perp}^\mathit{rms}}
\def\Rsparrms{R_\mathit{sol,\parallel}^\mathit{rms}}

\begin{document}

\title{
Self-bound quark matter in the NJL model revisited:\\
from schematic droplets to domain-wall solitons}

\author{Michael~Buballa}
\author{Stefano~Carignano}

\affiliation{Institut f\"ur Kernphysik (Theoriezentrum), Technische Universit\"at Darmstadt, Germany}

\date{February 2013}

\begin{abstract}
The existence and the properties of self-bound quark matter in the NJL model 
at zero temperature are investigated in mean-field approximation, 
focusing on inhomogeneous structures with one-dimensional spatial modulations.
It is found that the most stable homogeneous solutions which have previously
been interpreted as schematic quark droplets are unstable against formation of
a one-dimensional lattice of domain-wall solitons. The solitons repel each 
other,
so that the minimal energy per quark is realized in the single-soliton limit.
The properties of the solitons and their interactions are discussed in detail,
and the effect of vector interactions is estimated. 
The results may be relevant for the dynamics of expanding quark matter.
\end{abstract}

\maketitle

\section{Introduction}

The Nambu--Jona-Lasinio (NJL) model \cite{NJL} is a popular tool for 
studying low-energy properties of strongly interacting matter, like spectra
and scattering of light hadrons, or the phase diagram at nonvanishing
temperatures or densities (for reviews, see 
Refs.~\cite{Vogl:1991qt,Klevansky:1992qe,Hatsuda:1994pi,Buballa:2005}). 
While being relatively simple, the NJL model
shares the global symmetries of QCD, in particular chiral symmetry,
which is considered to be the most important feature of the model.
On the other hand, it is well known that the NJL model lacks confinement. 
In this sense it can be viewed as complementary to the MIT bag 
model~\cite{MIT},
which is confining by construction, but violates chiral symmetry at the
surface.

Some time ago, it was realized, however, that for sufficiently attractive 
interactions, the NJL model at zero temperature has solutions of self-bound  
chirally restored quark matter, which can be
interpreted as bag-model-like quark 
droplets~\cite{Buballa:2005,Buballa:1996tm,Alford:1997zt,Buballa:1998ky}.
In fact, the link between both models is
the existence of a ``bag pressure'', which in the bag model is introduced
by hand in order to stabilize the solutions, whereas 
in the NJL model it is a dynamical consequence of spontaneous chiral 
symmetry breaking in vacuum. 

The self-bound quark matter solutions mentioned above
have been obtained in the thermodynamic limit and correspond 
to infinite homogeneous matter. 
Their interpretation as quark-matter droplets is based on the behavior 
of the energy per particle, $E/N$, which shows a minimum at some 
nonvanishing saturation density.  
This means, a finite piece of quark matter with this density
would be stable against collapse or expansion, just like a liquid drop. 

An equivalent statement is that the matter has vanishing pressure, 
which means, it is in mechanical equilibrium with the vacuum. 
Thus, in order to have a solution of this type, there must be a phase 
coexistence of the vacuum with a dense-matter phase. In other words,
at some critical chemical potential, there must be a first-order phase 
transition from the vacuum to dense matter. 
This is realized in the NJL model, if the interaction is 
sufficiently attractive. 
On the other hand, if the attraction is relatively weak
(a condition which can be achieved, e.g., by adding a repulsive vector 
interaction),
it is also possible to have a second-order phase transition or a crossover.
In this case there is no stable matter solution and $E/N$ takes its
minimal value at zero density. 
Without applying external forces, a finite piece of quark matter would 
then keep expanding, i.e., behave like a gas. 

It is tempting at this point to extrapolate the self-bound solutions 
down to droplets 
consisting of only three quarks, and to interpret them as baryons.
However, although some of the resulting ``baryon'' properties are
quite reasonable~\cite{Buballa:1996tm}, it is obvious that
this extrapolation is not reliable. 
In fact, using solutions for infinite homogeneous quark matter to describe 
finite droplets, 
one has to assume that surface effects can be neglected.
This assumption might be justified for large droplets but most likely
not for small ones.
Besides, if the surface tension is positive, as derived, e.g., in 
Refs.~\cite{Molodtsov:2011ii,Lugones:2011xv,Pinto:2012aq}, 
smaller droplets are disfavored. 
The preferred state in the model would therefore be 
a configuration where all quarks are joined in one big spherical nugget, 
rather than hadronized into individual baryons.

It turns out, however, that this is not quite the case. More recent
studies of the NJL phase diagram have revealed that the first-order chiral
phase transition between homogeneous phases gets replaced by an 
inhomogeneous region if one allows the chiral condensate to be nonuniform 
in space~\cite{Sadzikowski:2000ap,Nakano:2004cd,Nickel:2009wj}. 
In this region, a special class of solutions, which vary in one spatial 
dimension, has been found to be favored over all other shapes considered so 
far. These solutions correspond to a lattice of domain-wall 
solitons\footnote{For brevity, we will just call them ``solitons''
in the following.}
described in terms of Jacobi elliptic functions and 
smoothly interpolate between the homogeneous chirally broken and restored 
phases~\cite{Nickel:2009wj}. 
In particular at the low-density side, they take the form of a single 
soliton, which is thermodynamically degenerate with the homogeneous chirally 
broken phase. As a consequence, the phase transition to the
latter is of second order.

This changes our picture of self-bound quark matter in the NJL model 
considerably. 
Since there is no longer a first-order phase transition connecting the
vacuum with a finite-density phase, but a second-order phase transition
to the inhomogeneous phase,  
the minimal $E/N$ should now be reached at zero average density.  
However, unlike the homogeneous case where the low-density regime 
corresponds to a dilute gas of constituent quarks, we now expect
a ``liquid crystal'' of well separated solitons. 
These objects have a nonvanishing quark density and a finite size in
one spatial dimension, while being infinite in the remaining two 
dimensions.
This could be seen as a step towards ``real'' quark droplets, which are
finite in three dimensions. 

In the present article we perform an explicit model study 
to investigate the scenario outlined above quantitatively.
After briefly introducing the formal background,
we calculate $E/N$ as a function of the average density
and compare the results for inhomogeneous solutions 
with those for homogeneous matter. 
Based on these results, we then discuss the properties of single
solitons and their interactions.
Finally, we estimate the effect of vector interactions, 
before we draw our conclusions.

\section{Nonuniform quark matter in the NJL model}
\label{sec:model}

In this section we briefly summarize the main properties of 
the one-dimensional solitonic NJL-model solutions derived in
Refs.~\cite{Nickel:2009wj} and \cite{Carignano:2010ac}.
Afterwards we study the single-soliton limit of these expressions.

\subsection{Mass functions and thermodynamic potential}

Our starting point is the Nambu-Jona Lasinio Lagrangian \cite{NJL} 
in the chiral limit,
\beq
  \mathcal{L}_{NJL} = \bar\psi i\gamma^\mu\partial_\mu \psi 
  + G \left( ({\bar\psi\psi})^2 + (\bar\psi i \gamma^5 \tau_a \psi)^2\right)\,,
\eeq
where $\psi$ denotes a quark field with $N_f=2$ flavor and $N_c=3$ color
degrees of freedom, $\tau_a$ are the Pauli matrices in flavor space,
and $G$ is a dimensionful coupling constant. 
The model is studied in the mean-field approximation.
To this end, we assume the presence of a nonvanishing scalar condensate, 
$\langle\bar{\psi}\psi\rangle=S(z)$,
which we allow to vary in one spatial dimension ($z$ direction) while 
being constant in the two perpendicular directions ($x$ and $y$) and in 
time.\footnote{
Other cases, like chiral density waves, which also include
pseudoscalar condensates \cite{Sadzikowski:2000ap,Nakano:2004cd},
or two-dimensional crystals \cite{Carignano:2012sx}
have been considered as well, but have been found to be less favored
at low densities~\cite{Nickel:2009wj,Abuki:2011pf,Carignano:2012sx}.
} 
Accordingly, the quarks acquire a $z$-dependent dynamical mass function
$M(z) = -2G \,S(z)$.

With this ansatz, one can employ the known results for the 
1+1-dimensional Gross-Neveu model \cite{Schnetz:2004vr}, 
to construct solutions of the 3+1-dimensional problem~\cite{Nickel:2009wj}.
The mass function can be expressed in terms of 
Jacobi elliptic functions,
\beq 
M(z) = \Delta\nu \frac{\text{sn}(\Delta z\vert \nu) 
\text{cn}(\Delta z\vert \nu)}{\text{dn}(\Delta z\vert \nu)} \,, 
\label{eq:mass}
\eeq
characterized by two parameters:
an amplitude $\Delta$ and the so-called elliptic modulus 
$\nu \in [0,1]$.
The latter determines the shape of the modulation, continuously changing
from sinusoidal for $\nu = 0$ to a hyperbolic tangent (``kink'')
for $\nu = 1$.
For $\nu < 1$, $M(z)$ is periodic with period~\cite{Schnetz:2004vr}
\beq
   L = \frac{2}{\Delta}\K(\nu)\,,
\label{eq:period}
\eeq
where $\K$ is the complete elliptic integral of 1st kind.

For the thermodynamic potential per volume at temperature  $T$ and
chemical potential $\mu$ one obtains
\bea
\label{eq:OmegaNJLNum}
\Omega(T,\mu;\Delta,\nu)
&=&
-
N_fN_c
\int_0^\infty\!dE\,
{\rho_\mathit{inh}}(E;\Delta,\nu)
\left[f_{vac}(E) + f_{med}(E;T,\mu) \right]
\nonumber\\
&+&
\frac{1}{4G_SL}\int_0^L
\!dz\,
\vert M(z)\vert^2
\,,
\eea
with the density of states
\bea
\label{eq:rho}
{\rho}_\mathit{inh}(E;\Delta,\nu)
=
\frac{E\Delta}{\pi^2}\left\{\phantom{\frac{\E}{\K}}\right.\hspace{-5mm}
&&
\theta(\sqrt{\tilde{\nu}}\Delta-E)
\left[
 \E(\tilde{\theta} \vert\tilde{\nu})+\left(\frac{\E(\nu)}{\K(\nu)}-1\right) 
\F(\tilde{\theta} \vert\tilde{\nu})
\right]
\nonumber\\
&+&
\theta(E-\sqrt{\tilde{\nu}}\Delta)
\theta(\Delta-E)
\left[
 \E(\tilde{\nu})+\left(\frac{\E(\nu)}{\K(\nu)}-1\right) \K(\tilde{\nu})
\right]
\nonumber\\
&+&
\left.
\theta(E-\Delta)
\left[
\E(\theta \vert\tilde{\nu})
+\left(\frac{\E(\nu)}{\K(\nu)} -1\right)\F(\theta \vert\tilde{\nu})
+\frac{\sqrt{(E^2 - \Delta^2) (E^2 - \tilde{\nu}\Delta^2)}}{E\Delta}
\right]
\right\}.
\nonumber\\
\eea
Here {\K} is again the complete elliptic integral of 1st kind,
{\F} is the incomplete elliptic integral of 1st kind, 
and {\E} are the (complete or incomplete) elliptic integrals of 2nd kind.
Furthermore we introduced the notations
$\tilde{\nu}=1-\nu$, $\tilde{\theta}=\arcsin(E/(\sqrt{\tilde{\nu}}\Delta))$, 
and $\theta=\arcsin(\Delta/E)$.

The functions ${f}_\mathit{vac}$ and ${f}_\mathit{med}(E)$ in
\eq{eq:OmegaNJLNum} are given by
\beq
{f}_{\mathit{vac}}(E) = E
\eeq
and 
\beq
{f}_\mathit{med}(E;T,\mu) =
T
\log\left(
1+\exp\left(-\frac{E-\mu}{T}\right)
\right)
+
T
\log\left(
1+\exp\left(-\frac{E+\mu}{T}\right)
\right)\,.
\eeq
Since the vacuum part of the energy integral is divergent,
we have to regularize it. 
We use Pauli-Villars regularization of the form~\cite{Klevansky:1992qe}
\bea
f_\mathit{vac}(E)
\;\rightarrow\;
\sum_{j=0}^{3}c_j\sqrt{E^2+j\Lambda^2}
\,,
\eea
with $c_0=1$, $c_1=-3$, $c_2=3$, $c_3=-1$ and a cutoff parameter $\Lambda$.

With these expressions at hand, the ground state of the system can be 
determined by minimizing the thermodynamic potential in the two parameters
$\Delta$ and $\nu$.

\subsection{Density profile}
\label{sec:dens}

The density profiles of the above 
solutions are given by~\cite{Carignano:2010ac}
\beq
n(z)
=
N_f N_c
\int\limits_0^\infty  dE\, \rho_{\mathit D,inh}(E,z;\Delta,\nu) 
\left(n_+(E) - n_-(E)\right)
\,,
\label{eq:density}
\eeq
where 
\beq
    n_\pm(E) = \frac{1}{e^{(E \mp \mu)/T} + 1}
\eeq
are the Fermi occupation functions for particles and antiparticles,
respectively,
and the density matrix element $\rho_\mathit{D,inh}$ can be related 
to $\rho_\mathit{inh}$, \eq{eq:rho}, upon the replacement
\beq
\rho_\mathit{D,inh}(E,z;\Delta,\nu)
=
\rho_\mathit{inh}(E;\Delta,\nu) \Big\vert_{
\frac{ {\mathbf E}(\nu )}{{\mathbf K}(\nu )} \rightarrow 
-\frac{1}{2} \left( \left(\frac{M(z)}{\Delta}\right)^2 + \nu -2 \right)
}
\,.
\label{eq:rho_D}
\eeq 

\begin{figure}
\begin{center}
\includegraphics[height=.45\textwidth,angle=-90]{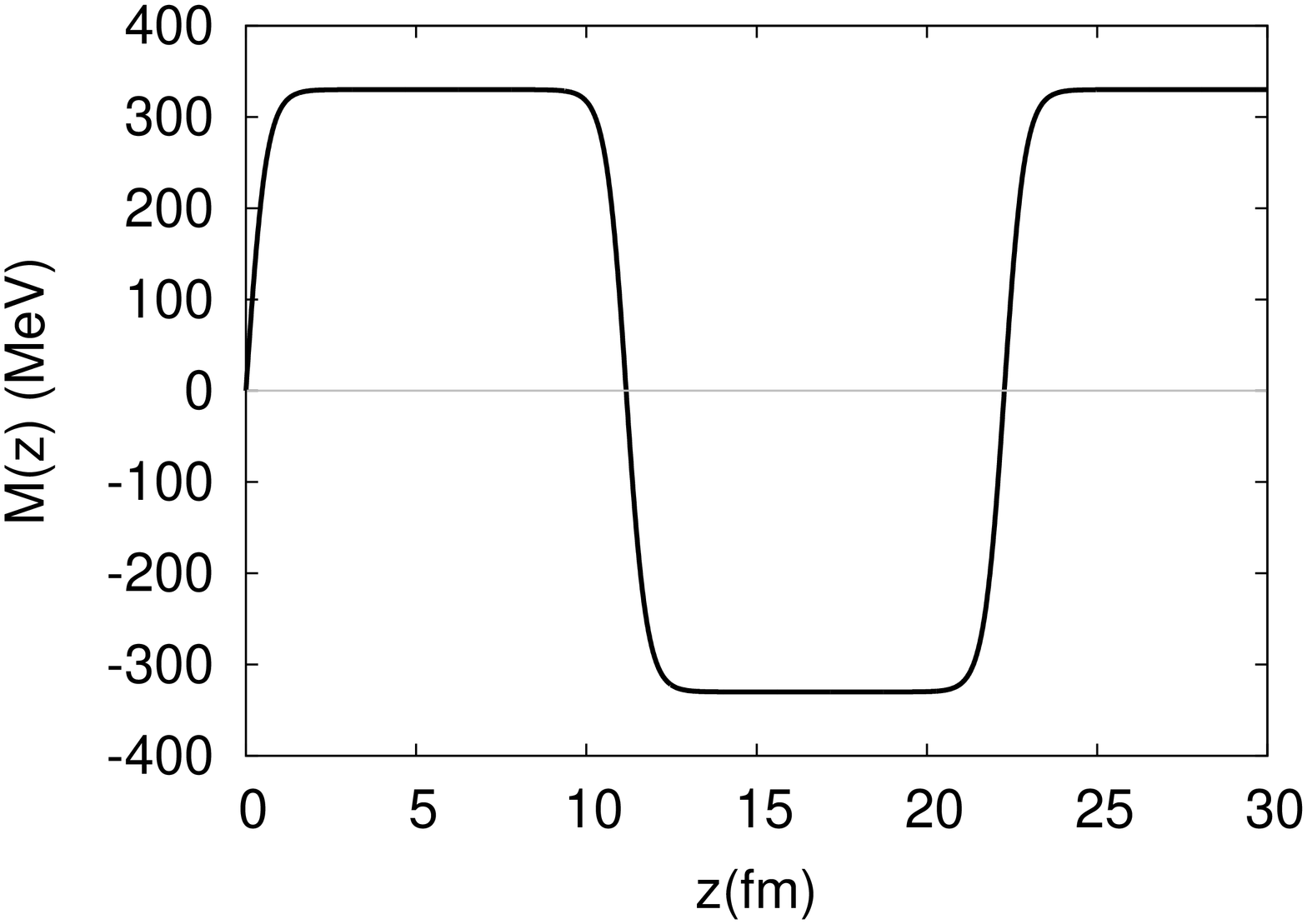}
\includegraphics[height=.45\textwidth,angle=-90]{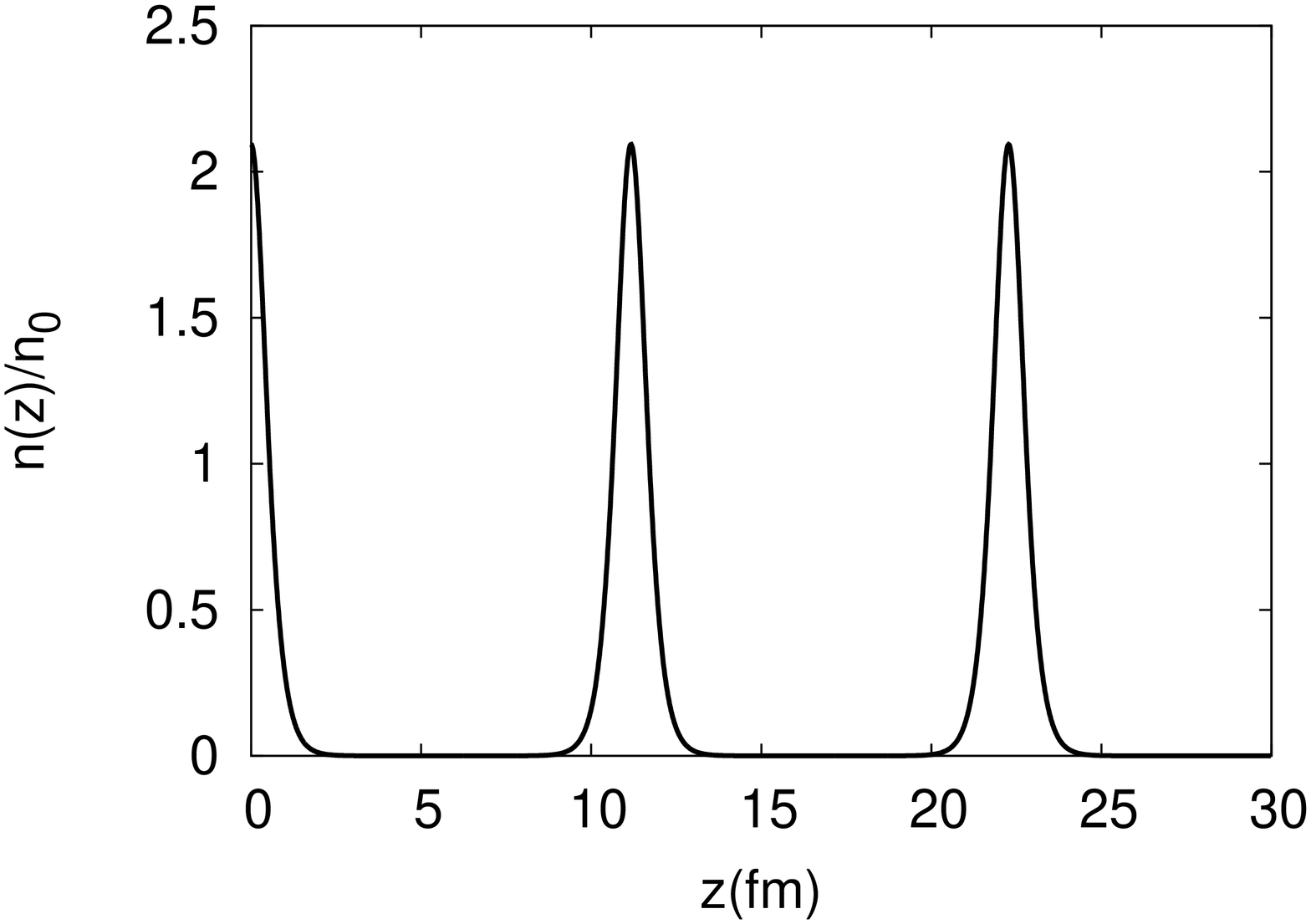}
\end{center}
\caption{Left: Mass function $M(z)$ for 
$\Delta = 330$~MeV and $\nu = 1 - 10^{-14}$. 
Right: Corresponding density profile at $T=0$ and $\mu = 323.3$~MeV. 
} 
\label{fig:M-n}
\end{figure}

As an example, we show in Fig.~\ref{fig:M-n} the mass function 
for $\Delta = 330$~MeV and $\nu = 1 - 10^{-14}$ (left),
and the corresponding density profile at $T=0$ and $\mu = 323.3$~MeV
(right). Comparing these figures, one can see that
the density is peaked at the points where the mass functions vanish, i.e., 
the regions of high density correspond to the regions where chiral
symmetry is almost restored. This is reminiscent of the bag model,
where the quarks are only allowed in the trivial vacuum.

Because of the alternating sign of the mass function, the density peaks
could be identified with solitons and antisolitons when projected onto
one spatial dimension parallel to the $z$-axis. In $3+1$ dimensions, 
this distinction is not well defined because the domain-walls can be 
oriented in any direction so that ``solitons'' and ``antisolitons'' are 
connected by a continuous transformation. 
In any case, as obvious from \eq{eq:rho_D}, the density does not depend 
on the sign of the mass function.
Therefore the distance $a$ between two neighboring peaks
is equal to one half of the period $L$,
where $L$ is given in \eq{eq:period}, 
\beq
    a = \frac{L}{2}
      = \frac{\K(\nu)}{\Delta}\,.
\label{eq:distance}
\eeq

\subsection{Single-soliton limit}
\label{sec:ssl}

In the limit $\nu \rightarrow 1$, the period $L$ goes to infinity,
and the mass function, \eq{eq:mass}, features a single kink at $z=0$,
\beq
    M(z)|_{\nu=1} = \Delta \, \tanh(\Delta z)\,,
\eeq
corresponding to a single soliton.
The density of states, \eq{eq:rho}, becomes
\beq
{\rho}_\mathit{inh}(E;\Delta,\nu=1)
\,=\,
\theta(E-\Delta)\,
\frac{1}{\pi^2} E \sqrt{E^2 - \Delta^2}
\,\equiv\, 
{\rho}_\mathit{hom}(E;\Delta)
\,,
\label{eq:rho_hom}
\eeq
which is equal to the density of states of an ideal gas of quarks 
with constant mass $\Delta$. 
As a consequence, the free energy of the inhomogeneous phase in the
single-soliton limit becomes degenerate with the free energy of
homogeneous matter with a constituent quark mass $\Delta$.

For the density matrix element \eq{eq:rho_D} one gets
\beq
\rho_\mathit{D,inh}(E,z;\Delta,\nu=1)
= 
\rho_\mathit{hom}(E;\Delta)
+
\rho_\mathit{D,sol}(E,z;\Delta)\,,
\eeq
with a localized part
\beq
\rho_\mathit{D,sol}(E,z;\Delta) =
\frac{E\Delta}{4\pi} \, \left(
\theta(\Delta-E) + \theta(E-\Delta) \frac{2}{\pi} \arcsin{\frac{\Delta}{E}}
\right)\,\frac{1}{\cosh^2(\Delta z)}
\eeq 
and a homogeneous background given by \eq{eq:rho_hom}.
The latter is again equal to the analogous term in homogeneous matter.

Accordingly, the density one obtains from \eq{eq:density} can be
separated into a constant background, which is equal to the density 
in a homogeneous ideal gas of quarks with mass $\Delta$
at given temperature and  chemical potential,
and a localized peak, which corresponds to the extra quarks in the solitons.
In particular, since the localized part drops off exponentially at
large values of $|z|$, the {\it average} density 
\beq
\nave = \lim_{L\rightarrow\infty} \frac{1}{2L}\int\limits_{-L}^{L} dz\,n(z)
\eeq
is entirely determined by the background and, thus, equal to the density
in homogeneous matter. 
As a consequence, a phase transition from the inhomogeneous phase to the 
homogeneous chirally broken phase taking place at $\nu=1$ is second order.

The additional density contribution due to the
quarks in the solitons,
\beq
n_\mathit{sol}(z) = 
N_f N_c
\int\limits_0^\infty  dE\, \rho_{\mathit D,sol}(E,z;\Delta) 
\left(n_+(E) - n_-(E)\right)
\,,
\eeq
is perhaps the most interesting part. 
In particular, $n_\mathit{sol}(z)$ and, thus, $n(z)$ is nonzero 
even at $T=0$ and $\mu < \Delta$, when the background density vanishes.
In this case, which corresponds to a single soliton embedded in the 
vacuum, one finds
\beq
    n_\mathit{sol}(z)|_{\{\Delta>\mu,T=0\}} 
    \,=\, \frac{N_f N_c}{8\pi} \frac{\Delta\mu^2}{\cosh^2(\Delta z)}\,.
\label{eq:nssl}
\eeq
Note, however, that here we have simply {\it assumed} that solutions with 
$\nu=1$ and $\Delta>\mu$ exist at zero temperature.
Of course, we have to check whether this comes out of the minimization of 
the thermodynamic potential. 
Since $\nu = 1$ is realized exactly at the second-order phase transition 
from the homogeneous to the inhomogeneous chirally broken phase, 
this means that at the critical chemical potential $\muci$, 
the amplitude $\Delta$ must be bigger than $\muci$. 
On the other hand, at $\mu = \muci$, the amplitude $\Delta$ is equal
to the constituent mass $M$ in the homogeneous phase. 
Moreover, in the homogeneous chirally broken phase, $M$ remains 
equal to the vacuum mass $M_\mathit{vac}$ as long as  
$\mu < M_\mathit{vac}$. Hence, if at $T=0$ there is a second-order phase 
transition to the inhomogeneous phase at $\muci < M_\mathit{vac}$,
then a single-soliton solution exists at $\mu = \muci$, 
with the density profile given by \eq{eq:nssl} and $\Delta = M_\mathit{vac}$.

In the next section this will be investigated further from the 
energy-per-particle perspective.

\section{Energy per particle}

Starting from the thermodynamic potential,
other thermodynamic quantities can be derived in the usual way,
as long as we are only interested in spatial averages. 
Restricting ourselves to zero temperature, this means that
the pressure $p$, the averaged quark number density $\nave$ and the 
averaged energy density $\epsave$ are given by
\beq
    p = -(\Omega(\mu) - \Omega_\mathit{vac})\,, \qquad
    \nave = \frac{\partial p}{\partial\mu}, \qquad
    \epsave = -p + \mu \nave\,.
\label{eq:p_n_eps}
\eeq
Here $\Omega_\mathit{vac}$ is the value of the thermodynamic potential
at its minimum in vacuum, which we subtract to define the vacuum pressure 
to be zero. 
As a consequence, the energy density of the vacuum vanishes as well. 
The average energy per quark is then given by
\beq
    \frac{E}{N} = \frac{\epsave}{\nave} = 
    -\frac{p}{\nave} + \mu\,.
\label{eq:EN}
\eeq
In the context of the interpretation of quark droplets as ``baryons'', 
the thermodynamics is often discussed in terms of the baryon number density 
$\rho_B = n/N_c$ and the energy per baryon $E/A = N_c E/N$, see, e.g., 
Refs.~\cite{Buballa:1996tm,Buballa:1998ky,Fukushima:2012mz}.
However, for most quantities we are going to discuss in this article,
it is more natural to work with quark number densities and $E/N$.
We therefore keep the notation introduced above, noting that the 
conversion to baryon quantities is simply a factor of $N_c = 3$.
Moreover, in our numerical examples we will scale the densities
by $n_0 = N_c \rho_0$, so that $n/n_0 = \rho_B/\rho_0$. 
Here $\rho_0 = 0.17$~fm$^{-3}$ is the nuclear matter saturation density,
i.e., the corresponding quark number density is $n_0 = 0.51$~fm$^{-3}$.

From \eq{eq:EN}, it follows that the density derivative of $E/N$ is given by
\beq
    \frac{\partial}{\partial\nave} \left(\frac{E}{N}\right)
    =
    \frac{1}{\nave}\left(\mu - \frac{E}{N}\right) 
    =
    \frac{p}{\nave^2}\,,
\label{eq:dENdn}
\eeq
where we have used that $\mu = \partial\epsave/\partial\nave$ at 
$T=0$ and fixed volume.
For $\nave \neq 0$, this means that $E/N$ has an extremum at the points
where the pressure vanishes, and it takes the value $E/N = \mu$ at 
these points. 

For $\nave \rightarrow 0$, on the other hand, the exact behavior 
of $E/N$ depends on the density dependence of the pressure. 
In the case of homogeneous quark matter, the NJL model at low densities 
behaves like an ideal nonrelativistic gas of constituent quarks, 
$p \propto n^{5/3}$. 
Consequently, $E/N$ goes to $\mu$,
which in turn converges to the vacuum constituent quark mass
$M_\mathit{vac}$,
while the density derivative of $E/N$ diverges at $n = 0$.
As we will see below, the behavior of inhomogeneous matter is rather
different.

\begin{figure}
\begin{center}
\includegraphics[width=.45\textwidth]{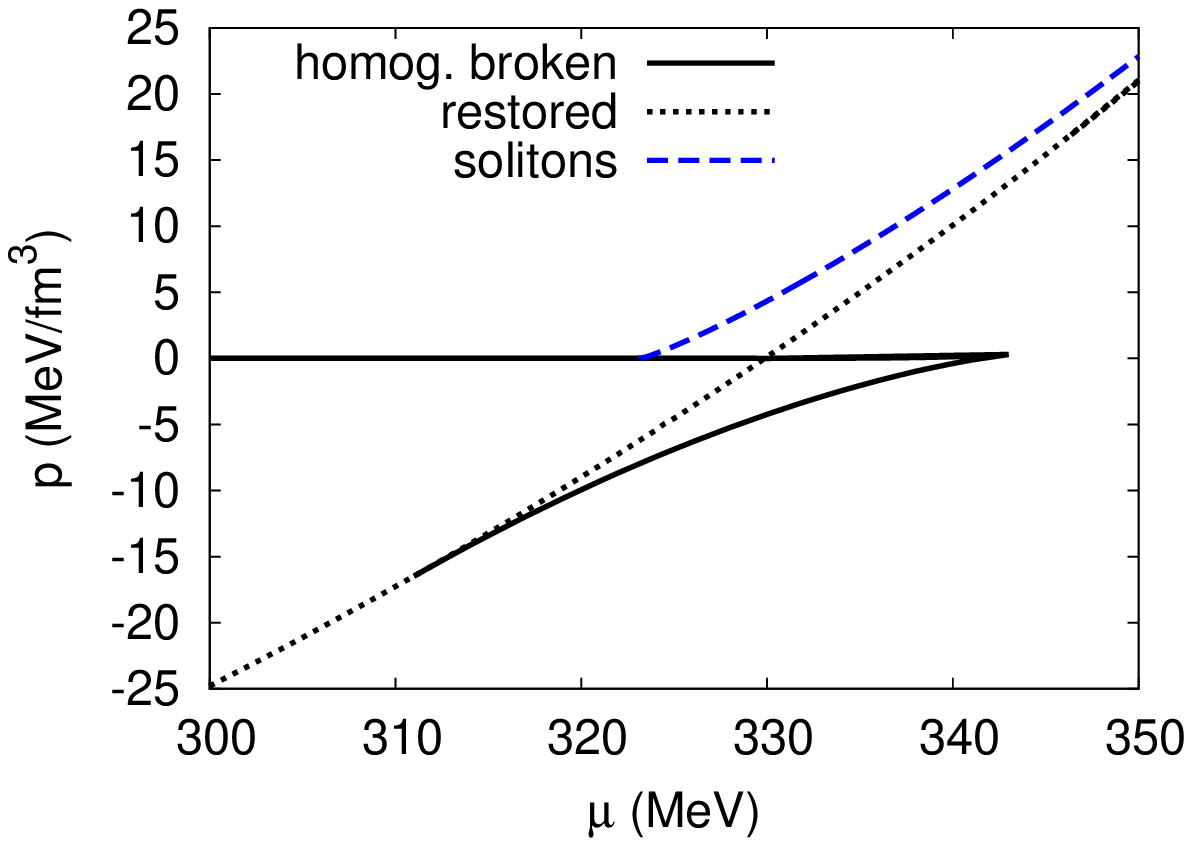}
\includegraphics[width=.45\textwidth]{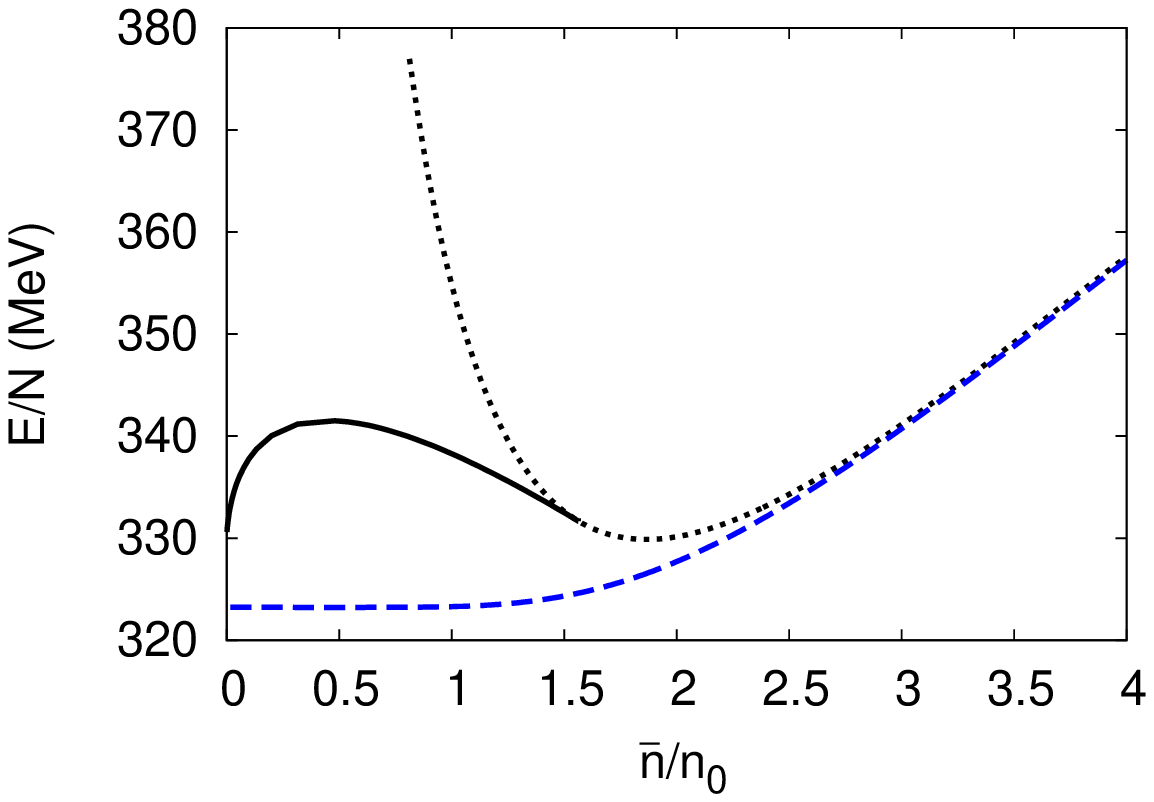}
\end{center}
\caption{Pressure as function of the chemical potential (left) 
and energy per quark as a function of the spatially averaged
quark number density (right).
The homogeneous chirally broken and restored solutions are indicated
by the solid and dotted lines, respectively, while the dashed lines
indicate the inhomogeneous solitonic solutions.} 
\label{fig:pmu_en}
\end{figure}

To this end, we now turn to the numerical results.
Our model contains two parameters: 
the coupling constant $G$ and the Pauli-Villars regulator $\Lambda$.
We fix them by fitting the pion decay constant in vacuum to its value
in the chiral limit, $f_\pi = 88$~MeV, and by choosing a reasonable 
value for the constituent quark mass in vacuum. 
If not stated otherwise, we choose $M_\mathit{vac} = 330$~MeV,
corresponding to $\Lambda = 728.368$~MeV and $G\Lambda^2 = 6.599$. 

In the left panel of Fig.~\ref{fig:pmu_en}, we show the pressure as
a function of $\mu$.
The homogeneous chirally broken solutions are indicated by the solid line, 
where the upper branch corresponds to the minima of the thermodynamic 
potential, i.e., to the stable or metastable solutions, while the lower
branch corresponds to the maxima, i.e., to the unstable solutions.
The chirally restored solutions are indicated by the dotted line. 
Restricting the analysis to these homogeneous
solutions, we find a first-order chiral phase transition at 
$\mu = \much = 329.9$~MeV, i.e., slightly below $\mu =M_\mathit{vac}$. 
Accordingly, the energy per particle in the restored phase, indicated 
by the dotted line in the right panel of  Fig.~\ref{fig:pmu_en}, has a minimum 
with $E/N = \much$ 
at $\nave = (2/\pi^2)\much^3$, 
whereas in the homogeneous chirally broken solution (solid line), $E/N$
is always larger, converging to $M_\mathit{vac}$ at $\nave = 0$
with an infinite slope. Thus, in the ``old picture'', 
we would interpret the minimum in the restored
phase as a bag-model-like quark droplet with a binding energy per quark of
$M_\mathit{vac} - \much$.

This picture is changed if we allow for the one-dimensional solitonic 
solutions, as indicated by the dashed lines in Fig.~\ref{fig:pmu_en}.
We then find a second-order phase transition\footnote{
This means that the elliptic modulus $\nu$ decreases continuously 
from $\nu = 1$ at $\muci$ to smaller values inside the inhomogeneous phase. 
Obviously, this is impossible to prove by numerical calculations.
Strictly speaking, we find that $\nu$ does not drop discontinuously from
1 to a value smaller than $1 - 10^{-14}$. At this point, the distance 
between the solitons is about $a=11$~fm, which is well above the size 
of the solitons (see Fig.~\ref{fig:M-n}). 
}
from the homogeneous chirally 
broken phase to the inhomogeneous phase at $\mu = \muci= 323.2$~MeV
(left panel). 
As discussed in Sect.~\ref{sec:ssl}, the inhomogeneous phase at this point
corresponds to a single soliton ($\nu = 1$) with vanishing background 
density.
Hence, there is no longer a stable solution with zero pressure and 
nonzero average density, and therefore the only minimum of the energy 
per particle exists at $\nave = 0$ (right panel).
On the other hand, the localized quarks inside the soliton experience
additional binding, so that $E/N$ does not go to $M_\mathit{vac}$ 
at $\nave = 0$, as for homogeneous matter, but to $\muci$, 
which is smaller than  $M_\mathit{vac}$ in this example.
The binding effect is also visible at nonvanishing $\nave$.
In particular, the chirally restored solution with the minimal $E/N$ 
is unstable against forming a soliton lattice with the same average 
density.
We find that here the solitons still have a sizable overlap, with
density peaks separated by about $a=1.5$~fm.
This system can then lower its energy further by expansion.

\begin{figure}
\begin{center}
\includegraphics[width=.45\textwidth]{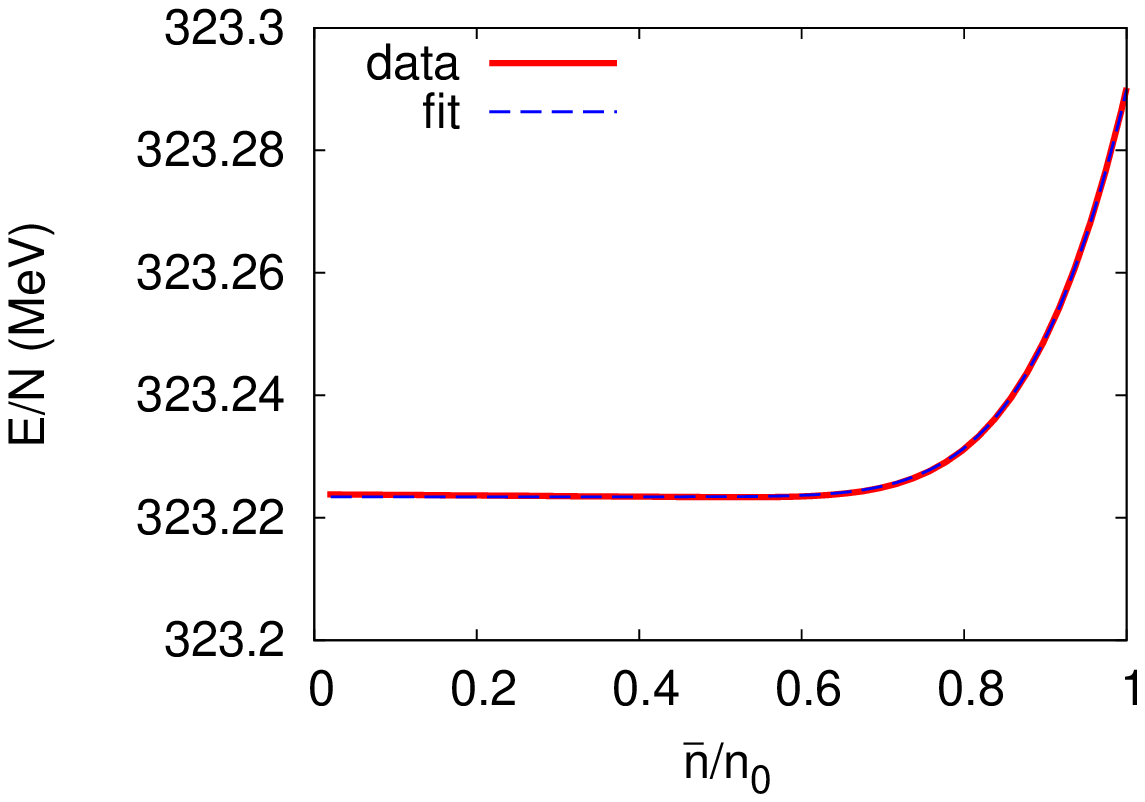}
\includegraphics[width=.45\textwidth]{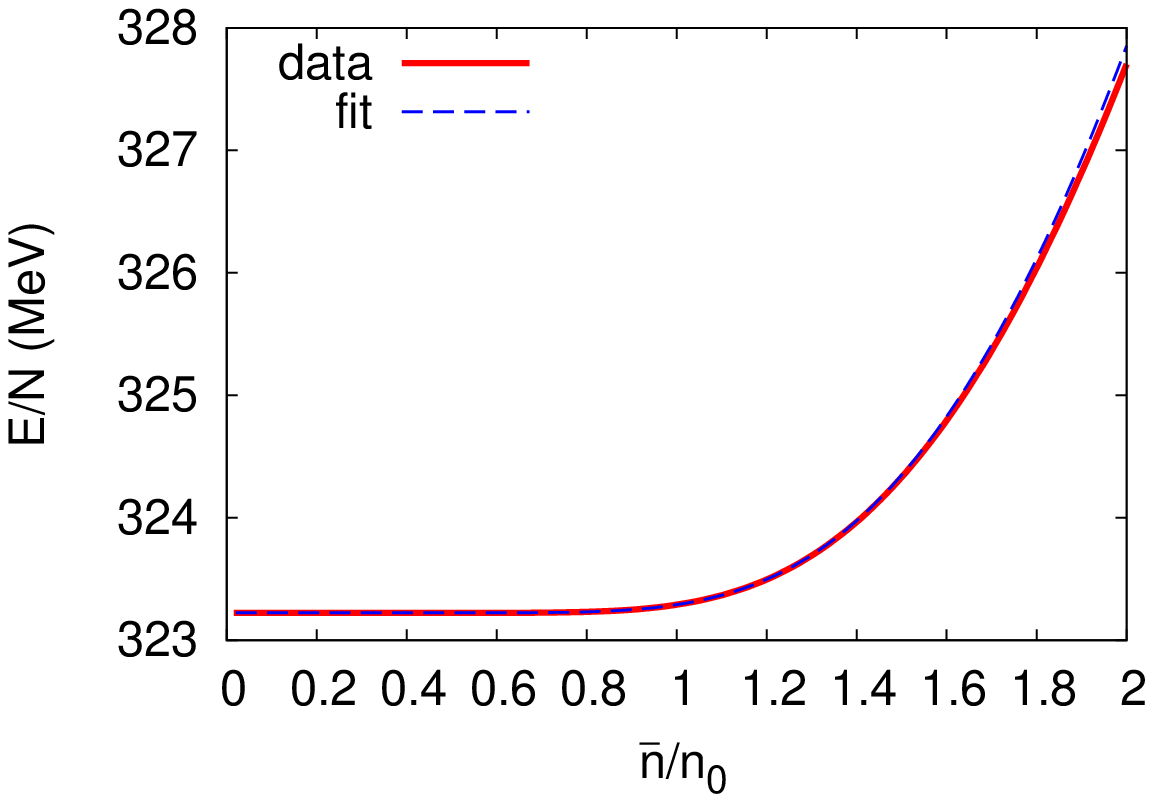}
\end{center}
\caption{Energy per particle as a function of the average density
at $T=0$: numerical results (red solid line) and according to 
\eq{eq:EoN_fit} with $c= 0.97985$ and $\muci = 323.223$ MeV 
(blue dashed line).
} 
\label{fig:EoN_fit}
\end{figure}

In this context, a
striking difference to the homogeneous case is the fact that 
the density derivative of $E/N$ does not diverge at $\nave = 0$
but, on the contrary, the function is extremely flat. 
According to \eq{eq:dENdn}, this means that the pressure goes to zero
with a high power of $n$.  
Further insight can be obtained from the observation in 
Ref.~\cite{Carignano:2010ac} that the density rise above the onset 
of the solitonic phase is consistent with the parametrization
\beq
\nave(\mu)
=
-\frac{c\muci^3}{\ln(\mu/\muci-1)}\,,
\label{eq:nave_mu}
\eeq
where $c$ is a constant parameter.
This formula was motivated by a similar behavior in the Gross-Neveu model. 
Strictly speaking, it describes the density change $\delta\nave$
relative to the density at $\mu =\muci$. 
However, since $\muci < M_\mathit{vac}$ in the present case, 
we have $\nave(\muci) = 0$ and, hence, $\delta\nave = \nave$.
We then find
\beq
    \frac{\partial p}{\partial \nave}
    = 
    \frac{\partial p}{\partial \mu} 
    \left(\frac{\partial \nave}{\partial \mu}\right)^{-1}
    =
    \frac{c \muci^4}{\nave}\, 
    e^{-c \muci^3/\nave}\,,
\label{eq:dpdn}
\eeq
where we have used \eqs{eq:p_n_eps} and
(\ref{eq:nave_mu}) to evaluate the derivatives.
It follows that
$\frac{\partial p}{\partial \nave}$ is exponentially suppressed
for  $\nave \rightarrow 0$.
The same is true for all higher derivatives 
and all derivatives of $E/N$, thus explaining its flatness.
In fact, integrating \eq{eq:dpdn} to obtain $p(\nave)$ and 
inverting \eq{eq:nave_mu} for $\mu(\nave)$, we get from 
\eq{eq:EN} that the energy per particle at low densities should be
given by
\beq
     \frac{E}{N} 
     = \muci \left( 1 + e^{-c\muci^3/\nave}
     -  \frac{c\muci^3}{\nave} \int\limits_{c\muci^3/\nave}^\infty
     \!\!\!dx \;\frac{e^{-x}}{x}\right)\,.
\label{eq:EoN_fit}
\eeq
In Fig.~\ref{fig:EoN_fit}, this expression is compared with the 
numerical results for $E/N$. Fitting the parameters $c$ and $\mu_{c,inh}$ 
to the data below $\nave = n_0$ (left), 
we find a reasonable description up $\nave = 2n_0$ (right),
where the increase of $E/N$ is more than a factor of 50 larger. 
We remark that the fitted value for $c$ is very close to 
$\muci/M_\mathit{vac}$, but we have not been able to show this analytically.

For completeness, we also comment on the behavior at high densities. 
In our example with $M_\mathit{vac} = 330$~MeV, the system stays
inhomogeneous up to arbitrarily high chemical potentials.
For somewhat lower values of $M_\mathit{vac}$, there is first a second-order 
phase transition from the solitonic phase to the restored phase, but the 
system gets inhomogeneous again at higher chemical potentials. 
As discussed in detail in Ref.~\cite{Carignano:2011gr}, 
this so-called ``inhomogeneous continent'' is not a trivial regularization
effect, but it cannot be excluded that it is a model artifact.
In this article, however, we are mainly interested in the low-density
behavior of the model, where this issue is irrelevant.

\begin{figure}
\begin{center}
  \includegraphics[width=.48\textwidth]{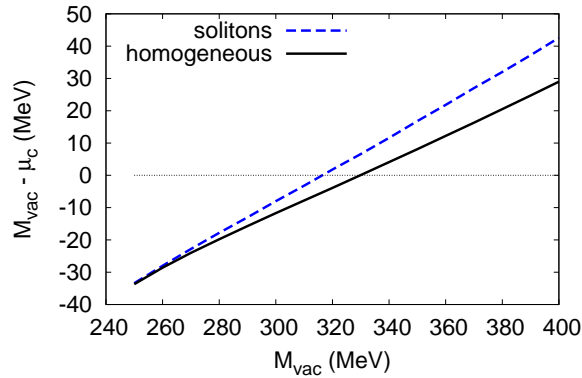}
\end{center}
\caption{Binding energy per quark for homogeneous matter (solid line)
and for solitons (dashed line) as functions of the 
vacuum constituent quark mass.} 
\label{fig:Ebind}
\end{figure}

The results shown in Fig.~\ref{fig:pmu_en} have been obtained for 
specific parameters, and one might wonder how robust they are when 
these are changed.
As discussed in Refs.~\cite{Buballa:2005,Buballa:1996tm,Buballa:1998ky},
the binding energy of homogeneous quark matter depends on the 
amount of scalar attraction, which can be parametrized by $M_\mathit{vac}$.
In Fig.~\ref{fig:Ebind}, we therefore show the binding energies per
quark in homogeneous matter and in the solitons,
$E_\mathit{b,hom}= M_\mathit{vac}-\much$
and $E_\mathit{b,sol}= M_\mathit{vac}-\muci$, respectively,
as functions of $M_\mathit{vac}$.

We see that both curves start at the same point around 
$M_\mathit{vac} = 250$~MeV with a negative binding energy.
For homogeneous matter, this point corresponds to the limiting case
where the phase transition turns from first to second order when
$M_\mathit{vac}$ is lowered further. In other words, this point corresponds 
to the case where the tricritical point of the phase boundary in the 
$T-\mu$ plane is just located at the $T=0$ axis. 
Since in this model the tricritical point is equal to the Lifshitz 
point~\cite{Nickel:2009ke}, 
i.e., the point where the two homogeneous phases and the inhomogeneous 
phase meet, the binding energies of homogeneous and 
inhomogeneous matter are equal at this point, and both solutions cease to 
exist at lower values of $M_\mathit{vac}$. 

When $M_\mathit{vac}$ is increased, the binding
energies rise. 
The  would-be first-order phase transition from the 
homogeneous chirally broken to the restored phase is now inside the
inhomogeneous regime, i.e., $\muci<\much$
and, hence, $E_\mathit{b,sol} > E_\mathit{b,hom}$.
This means, the chirally restored solution with the lowest $E/N$ is 
always unstable against forming a soliton lattice.

On the other hand, for $M_\mathit{vac} < 315$~MeV, 
$\muci$ is still smaller than $M_\mathit{vac}$.
Then the density in the homogeneous chirally broken phase is
already nonzero when the phase transition to the inhomogeneous phase takes
place. 
Thus, as discussed in Sec.~\ref{sec:ssl}, the soliton is embedded
in a homogeneous background of constituent quarks at this point.  
As the pressure is nonzero, the system wants to expand. 
However, since all solutions with a lower average density are homogeneous,
this means that the inhomogeneous phase, including
the single-soliton solution is not stable without applying 
external forces.\footnote{We have seen that inhomogeneous solutions with 
$\nu = 1$ and $\Delta = M$ are thermodynamically degenerate with 
homogeneous matter with mass $M$. Hence, one may argue that the single 
solitons survive also below $\mu = \muci$, down to  $\mu = M_\mathit{vac}$,
where they would have vanishing binding energy. 
On the other hand, they are no longer solutions of the gap equation 
$\frac{\partial\Omega}{\partial\nu}=0$, and it is therefore unclear
whether they are self-consistent and thermodynamically consistent solutions. 
Here we choose not to further investigate this question,
since in this work we are mainly interested in solutions with a 
positive binding energy, $\muci <  M_\mathit{vac}$. 
}
As a consequence, the lowest $E/N$ is obtained for a dilute
gas of constituent quarks in the limit of zero density.

For $M_\mathit{vac} > 315$~MeV, $E_\mathit{b,sol}$
is positive, i.e., the lowest $E/N$ corresponds to a single soliton state, 
as discussed above.
For $M_\mathit{vac} > 330$~MeV, $E_\mathit{b,hom}$ is positive as well.
So there would be stable droplets of homogeneous matter if we could ignore
the possibility of inhomogeneity.   
As explained above, this is, however, not the case. 
At $M_\mathit{vac} = 400$~MeV, for example, the binding energy for
homogeneous matter is about 30~MeV per quark, while it is about
40~MeV per quark for the solitons. 

For simplicity, all calculations in this article are done in the chiral 
limit. 
The mass functions and phase diagrams for non-vanishing bare quark
masses have been investigated in Refs.~\cite{Nickel:2009wj,Carignano:2010ac},
and turned out not to be very different.
In particular, the inhomogeneous phase is delimited by second-order
phase boundaries, 
and the mass function at the boundary towards lower $\mu$
takes the form of a single soliton. 
Therefore, we do not expect our results to change qualitatively if
finite bare quark masses are considered.

\section{Properties of single solitons}
\label{sec:pss}

After having explored the conditions for the existence of 
single self-bound solitons,
we would now like to investigate their properties in more details. 

As discussed in Sec.~\ref{sec:ssl}, the density profile is given by
\eq{eq:nssl} with $\mu = \muci$ and $\Delta = M_\mathit{vac}$,
\beq
    n_\mathit{sol}(z)
    \,=\, \frac{N_f N_c}{8\pi} \frac{M_\mathit{vac}\,\muci^2}
                                    {\cosh^2(M_\mathit{vac}\, z)}\,,
\label{eq:nssl2}
\eeq
where $\muci < M_\mathit{vac}$.
We have already noted that
$n_\mathit{sol}(z)$ decreases exponentially at large $|z|$
and therefore the average density $\nave$ vanishes.
On the other hand, the central density at $z=0$ 
is larger than the density of restored quark matter at the same
chemical potential, 
\beq
    n_\mathit{sol}(0) 
    \,=\, 
    \frac{N_f N_c}{8\pi} \, M_\mathit{vac}\,\muci^2
    \,=\, 
    \frac{3\pi}{8} \frac{M_\mathit{vac}}{\muci}\,n_\mathit{rest} \,,
\eeq
where
$n_\mathit{rest} = \frac{N_f N_c}{3\pi^2}\,\muci^3$.
This can be interpreted as a bag-pressure effect, which pushes the quarks 
out of the chirally broken vacuum and squeezes them into the restored 
regions~\cite{Carignano:2010ac}. 

The number of quarks in the soliton per transverse area $A_\perp$
is obtained by integrating \eq{eq:nssl2} over $z$.
One finds
\beq
    \frac{N}{A_\perp} \,=\, \frac{N_f N_c}{4\pi}\,\muci^2\,.
\label{eq:NoverA}
\eeq 
Another interesting quantity is the longitudinal rms ``radius'',
\beq
    \Rsparrms
    \,\equiv\,
    \sqrt{\ave{z^2}} 
    \,=\,
    \left(\frac{\int dz\, z^2\, n_\mathit{sol}(z)}
               {\int dz\, n_\mathit{sol}(z)} \right)^{1/2} 
    \,=\,
    \frac{\pi}{\sqrt{12}}\,\frac{1}{M_\mathit{vac}}\,,
\label{eq:deltaz}
\eeq
which is a measure for the half-size of the soliton in $z$-direction. 

Similarly, we can define the ``soliton averaged density'', i.e., 
the density-weighted integral over the density divided by the number of 
quarks,
\beq
    \ntyp 
    \,=\,
    \frac{\int dz\, n_\mathit{sol}^2(z)}{\int dz\, n_\mathit{sol}(z)}
    \,=\,
    \frac{N_f N_c}{12\pi}\,M_\mathit{vac}\,\muci^2\,.
    \label{eq:ntyp}
\eeq
Hence
\beq
    \ntyp
    \,=\,
    \frac{2}{3} n_\mathit{sol}(0)
    \,=\, 
    \frac{\pi}{4} \frac{M_\mathit{vac}}{\muci}\,n_\mathit{rest} \,,
\eeq
i.e., while the maximal density in a self-bound soliton is always
larger than the density in the restored phase at the same chemical
potential, $\ntyp$ can be smaller.
For instance, for $M_\mathit{vac} = 400$~MeV, we have 
$M_\mathit{vac}/\muci = 1.12 < 4/\pi$.

It is interesting to compare these expressions with the results based
on the droplet picture for homogeneous matter. 
As discussed earlier, the most stable homogeneous solution corresponds 
to quark matter in the restored phase at the critical chemical potential 
$\much$, provided $\much< M_\mathit{vac}$.
The density is, thus, given by
\beq
    n_\mathit{hom} = \frac{N_f N_c}{3\pi^2}\,\much^3\,.
\eeq
Assuming that the homogeneous solutions could be taken over to describe
small quark matter droplets, the volume of a ``baryon'' with $N_c$ quarks
would be $V_\mathit{hom} = N_c / n_\mathit{hom}$. 
For spherical bags, this would correspond to a radius of 
\beq
    R_\mathit{hom,s} 
    = \left(\frac{9\pi}{4 N_f}\right)^{1/3}\,\frac{1}{\much}\,,
\eeq
which turns out to be quite reasonable if the numerical values for
$\much$ are inserted~\cite{Buballa:1996tm}, see Table~\ref{tab:hom}.
However, since the underlying formalism, which assumes infinite matter,
does not provide any mechanism why the matter should clusterize 
and, if so, why into droplets of $N_c$ quarks,
this description of baryons as quark droplets remains very schematic. 

In this sense, the domain-wall solitons, which are finite  
in one spatial direction, could be seen as a step into the right direction.
Moreover, the longitudinal size, given by \eq{eq:deltaz}, turns out to 
be of the correct order. 
To illustrate this, we perform a quantitative comparison with 
the homogeneous ``baryon'' droplets 
by again restricting the volume in such a way that it contains $N_c$ quarks. 
Since the longitudinal shape of the soliton
is predicted by the model, we want to 
keep it untouched and only restrict the transverse area by hand.
Taking a circular shape, \eq{eq:NoverA} yields
\beq
    \Rsperp = \frac{2}{\sqrt{N_f}}\,\frac{1}{\muci}\,.
\eeq
Of course, the non-spherical geometry of this ``baryon'' should
not be taken seriously, but is simply a consequence of the
described procedure.
For the sake of a meaningful comparison with the homogeneous droplets,
we take the latter to be non-spherical as well, but assume a 
cylindrical shape.
For simplicity, we assume that the transverse and longitudinal radii
of the cylinder are equal, i.e., 
the cylinder has a transverse radius 
$R_\mathit{hom,c}$ and a height $2R_\mathit{hom,c}$. 
Since the volume must remain unchanged,
$R_\mathit{hom,c}$ is then related to the radius of the sphere by
$R_\mathit{hom,c} = (2/3)^{1/3} R_\mathit{hom,s}$, i.e.,
\beq
    R_\mathit{hom,c} 
    = \left(\frac{3\pi}{2 N_f}\right)^{1/3}\,\frac{1}{\much}\,.
\eeq
Moreover, for better comparability with $\Rsparrms$, 
we translate all sharp radii into rms radii.
For a D-dimensional sphere with radius $R_i$, the relation is given by
\beq
    R_i^\mathit{rms} = \left(
    \frac{\int d^D r\, r^2\, \theta(R_i -r)}
    {\int d^D r\, \theta(R_i-r)} \right)^{1/2}
    = \sqrt{\frac{D}{D+2}}\,R_i\,,
\eeq
i.e., we have 
$\Rsperprms/\Rsperp = R_\mathit{hom,\perp}^\mathit{rms}/R_\mathit{hom,c} 
= 1/\sqrt{2}$
and 
$R_\mathit{hom,\parallel}^\mathit{rms}/R_\mathit{hom,c} = 1/\sqrt{3}$.

Finally, we define a ``baryon mass'' for the homogeneous
and solitonic solutions as $N_c$ times $E/N$, i.e.,
\beq
    M_\mathit{B,hom} = N_c\, \much\,, \qquad 
    M_\mathit{B,sol} = N_c\, \muci\,.
\eeq

\begin{table*}
\begin{tabular}{|c|cccccc|}
\hline
\; $M_\mathit{vac}$ [MeV] \; & \; $\much$ [MeV] \; & 
\; $M_\mathit{B,hom}$ [MeV] \; & \; $n_\mathit{hom}/n_0$ \; &
\; $R_\mathit{hom,s}$ [fm] \; & 
\; $R_\mathit{hom,\perp}^\mathit{rms}$ [fm] \; & 
\; $R_\mathit{hom,\parallel}^\mathit{rms}$ [fm] \;
\\
\hline
330  &  329.9  &  \phantom{1}989.7  &  1.86  &  0.91  &  0.56  &  0.46
\\
400  & 371.0  &              1113.0 &  2.64  &  0.81  &  0.50  &  0.41
\\
\hline
\end{tabular}
\caption{Properties of homogeneous ``baryon droplets'' for two
values of the vacuum constituent quark mass $M_\mathit{vac}$. 
}
\label{tab:hom}
\end{table*}


\begin{table*}
\begin{tabular}{|c|ccccccc|}
\hline
\, $M_\mathit{vac}$ [MeV] \, & \, $\muci$ [MeV] \, & 
\, $M_\mathit{B,sol}$ [MeV] \, & \, $n_\mathit{sol}(0)/n_0$ \, &
\, $\ntyp/n_0$ \, &  \, $\Rsperp$ [fm] \, & 
\, $\Rsperprms$ [fm] \, & 
\, $\Rsparrms$ [fm] \,

\\
\hline
330  &  323.2  & \phantom{1}969.7  &  2.10  &  1.40  &  0.86  &  0.61  &  0.54
\\
400  &  357.4  &           1072.2  &  3.11  &  2.08  &  0.78   & 0.55  &  0.45
\\
\hline
\end{tabular}
\caption{Properties of solitonic ``baryons'' for two
values of the vacuum constituent quark mass $M_\mathit{vac}$. 
}
\label{tab:sol}
\end{table*}

The results for $N_f=2$ and $N_c=3$ and two different values of 
$M_\mathit{vac}$ are summarized in 
Table~\ref{tab:hom} for homogeneous droplets and in Table~\ref{tab:sol}
for the solitons.
We find that the qualitative and even the quantitative behavior is
similar for both cases.
The ``baryon masses'' rise with the vacuum quark mass, but are below
$3 M_\mathit{vac}$ because of binding effects. 
Since the binding increases with increasing quark masses, the densities
increase as well, while the radii decrease.
As discussed in the previous section, the solitons are bound more strongly
than homogeneous matter. As a consequence, the solitons have a larger
central density, despite the fact that the chemical potential is lower.
The soliton averaged density $\ntyp$, on the other hand, 
is smaller than the density in homogeneous droplets. 
Therefore, the solitons have larger rms radii.

Nevertheless, the general agreement of the various rms radii 
for a given quark mass turns out to be quite good.
In particular, it is remarkable that $\Rsparrms$,
which is an intrinsic property of the soliton, is similar to the
other radii, which have been introduced by hand in order to have three
quarks in a ``baryon''.
On the other hand, the sharp radii $R_\mathit{hom,s}$ and $\Rsperp$
are considerably larger, showing that the numbers are rather sensitive to
the used definition of the radius.

\section{Soliton-soliton interactions}
\label{sec:sasi}

Having discussed the properties of single solitons,
we now move away from this limit and investigate what happens when
the solitons approach each other. 

As explained in Sec.~\ref{sec:model}, the inhomogeneous solutions
are characterized by the parameters $\nu$ and $\Delta$, which are
obtained by minimizing the thermodynamic potential at given $T$ and 
$\mu$. In particular the distance $a$ between the neighboring solitons 
depends on $\nu$ and $\Delta$, as detailed in 
\eq{eq:distance}.
This allows us to plot the thermodynamic quantities of the system
as functions of $a$, which is sometimes more instructive than plotting 
them against $\mu$ or $\nave$.  

\begin{figure}
\begin{center}
\includegraphics[width=.45\textwidth]{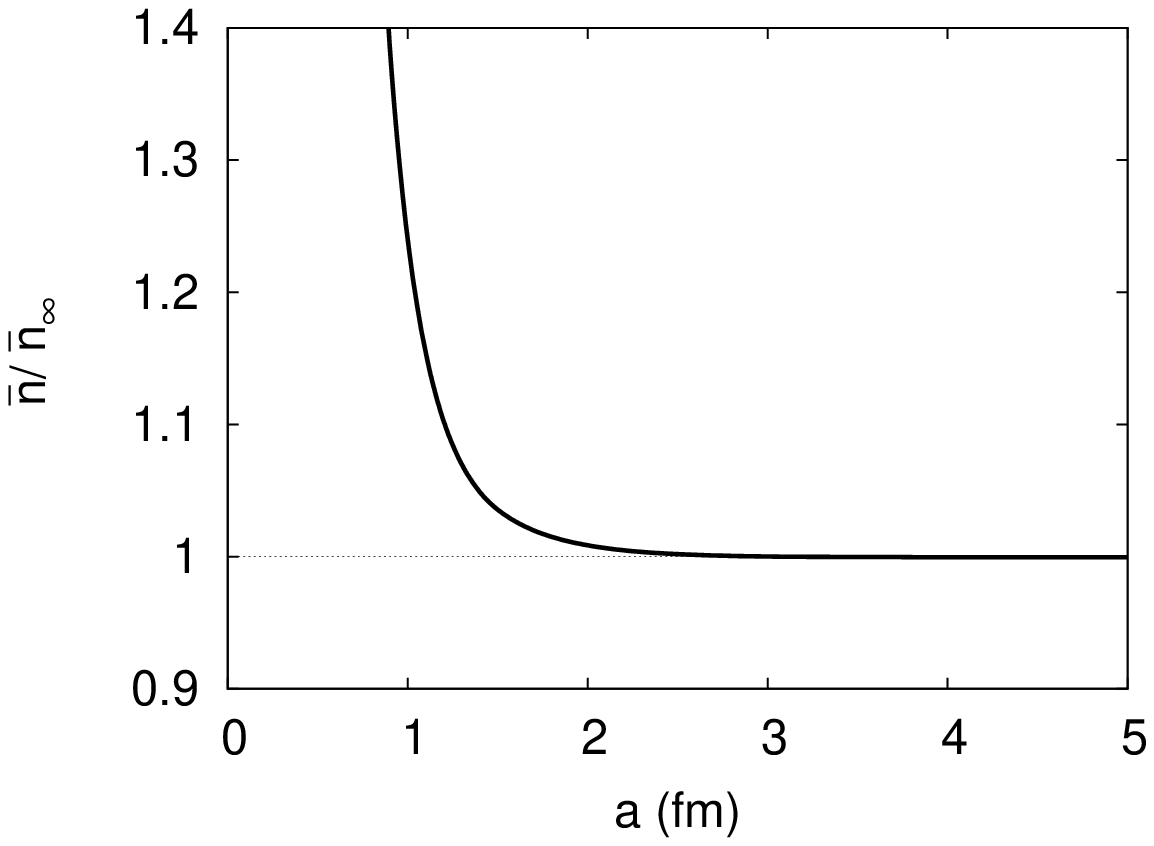}
\includegraphics[width=.45\textwidth]{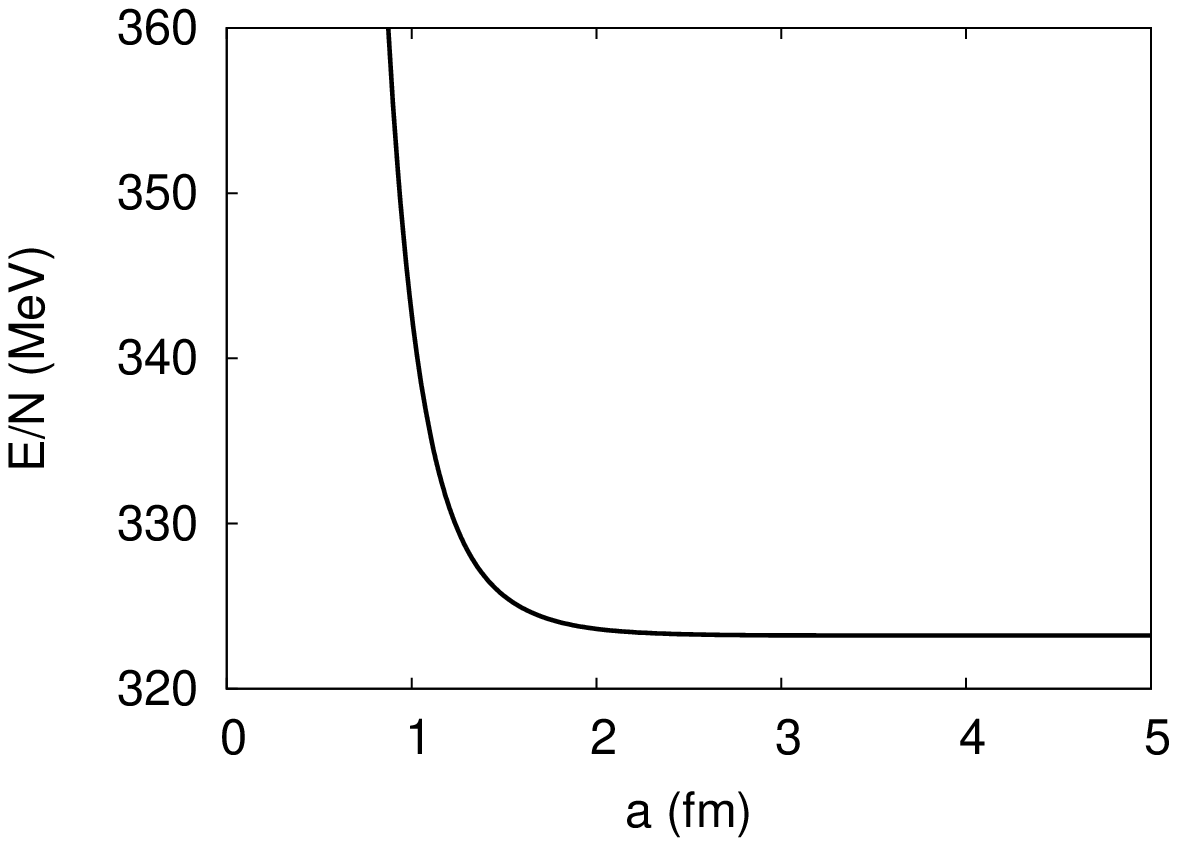}
\includegraphics[width=.45\textwidth]{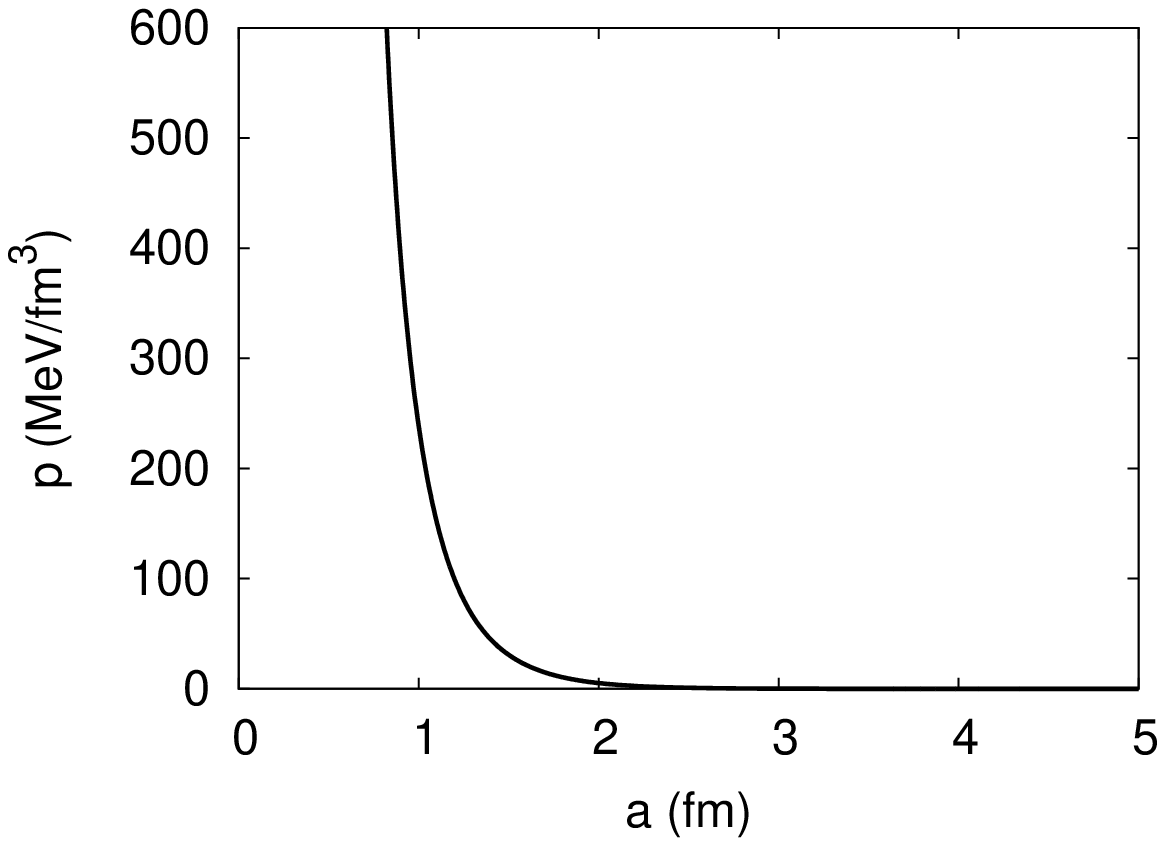}
\includegraphics[width=.45\textwidth]{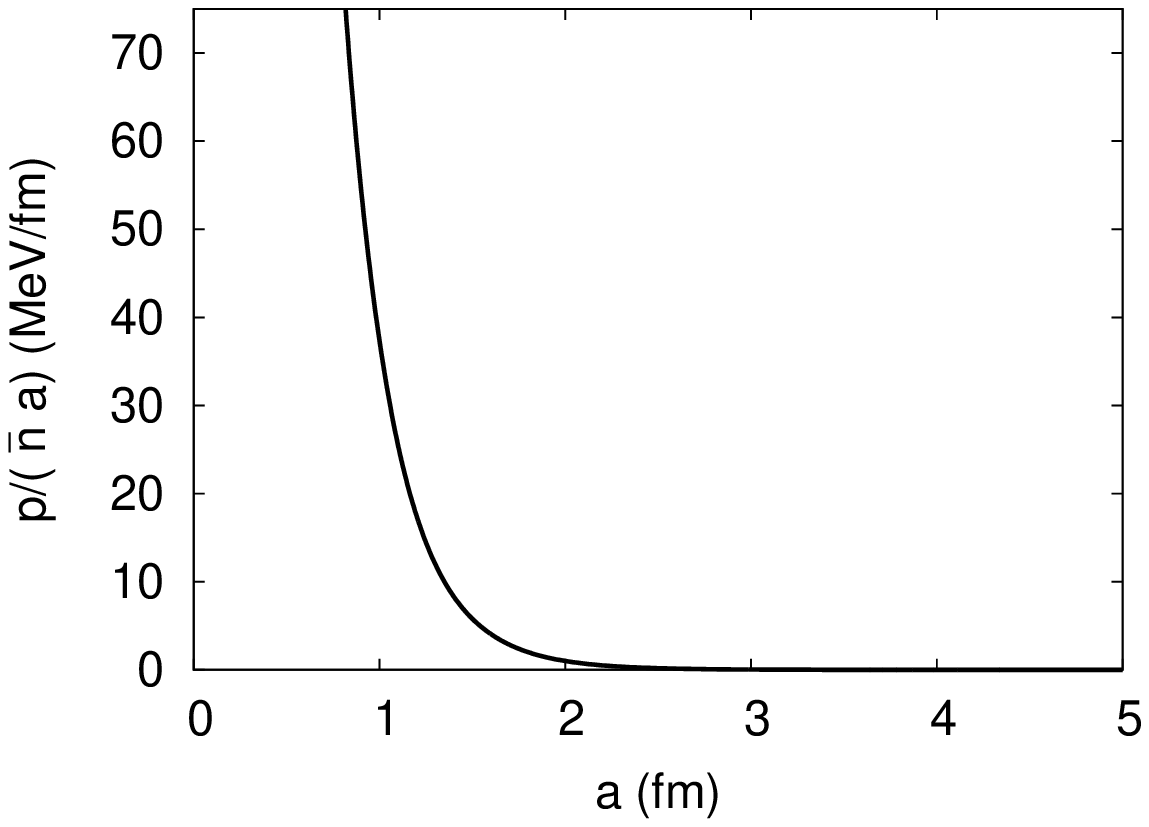}

\end{center}
\caption{Various quantities as functions of the soliton-soliton distance $a$:
ratio of the average density $\nave$ and the corresponding value 
$\nave_\infty$ for noninteracting solitons, \eq{eq:ninfty} (upper left),
energy per quark (upper right), pressure (lower left), 
and effective force per quark, \eq{eq:FoverN} (lower right).} 
\label{fig:nEoNFvsa}
\end{figure}

As before, we limit ourselves to $T=0$.
At the boundary to the homogeneous chirally broken phase, we have
$\nu=1$, corresponding to $a \rightarrow \infty$,
while with increasing chemical potential the distance quickly becomes
smaller.  
For large distances, the density distribution of the
soliton lattice does not differ much from a linear superposition 
of single solitons. 
The average density is therefore given by 
\beq
    \nave_\infty(a) = \frac{N_f N_c}{4\pi}\,\frac{\muci^2}{a}\,,
\label{eq:ninfty}
\eeq
i.e., the column density of a single soliton, \eq{eq:NoverA},
divided by the distance.
At smaller distances, on the other hand, the interaction between the 
solitons leads to nonlinearities, giving rise to deviations
from the trivial $1/a$ behavior. 
This is shown in the upper left panel of Fig.~\ref{fig:nEoNFvsa}, 
where $\nave/\nave_\infty$ is displayed as a function of $a$.
We see that the ratio is very close to unity for $a > 2$~fm
and rises steeply when the distance is decreased below 1~fm.
Here $a$ is smaller than $2\Rsparrms$, i.e., the
solitons strongly overlap.
A similar picture arises from the energy per particle (upper right)
and the pressure (lower left) when plotted as functions of the soliton 
distance: For $a > 2$~fm, $E/N$ is almost independent of $a$ and $p$
remains close to zero, while both quantities rise steeply at $a < 1$~fm.

For thin, well separated solitons, the pressure can be 
interpreted as the force per transverse area by which they repel each other, 
$F(A_\perp) = p A_\perp$.
Dividing this force by the corresponding number of quarks in the soliton,
$N(A_\perp) = \nave a A_\perp$,
we obtain the effective force per quark
\beq
\ave{\frac{F}{N}} = \frac{p}{\nave a}\,,
\label{eq:FoverN}
\eeq
which is probably the most intuitive way to quantify the 
soliton-soliton interactions.
The resulting behavior as a function of $a$ is shown in the lower right 
panel of Fig.~\ref{fig:nEoNFvsa}.
Again, the ``force'' vanishes at large distances and becomes 
nonnegligible only below around 2~fm, when the solitons begin to overlap.
Of course, when the overlap gets sizable, the assumption of well separated
solitons breaks down and the interpretation as a force must be taken with
care.

\section{Including vector interactions}
\label{sec:vec}

It is also interesting to study the influence of vector interactions, 
which are very important at finite density, as known, e.g., from the Walecka 
model~\cite{Walecka:1974qa}.
In the NJL model with homogeneous condensates, vector interactions have
been shown to weaken the first-order chiral phase 
transition, and already at rather small values of the vector coupling, 
the phase transition turns into second order or a crossover 
\cite{Buballa:1998ky, Kitazawa:2002bc,Fukushima:2008is,Bratovic:2012qs}. 
In terms of $E/N$, this is easily understood from the fact that the 
vector interaction, described by a term 
\beq
 {\cal L}_V = -G_V(\bar\psi\gamma^\mu\psi)^2
\eeq 
in the Lagrangian, adds a term $G_V n^2$ to the energy density, 
i.e., $E/N$ is enhanced by $G_V n$~\cite{Buballa:1996tm}. 
Hence, the minimum in the restored phase at finite density gets increasingly
disfavored with increasing $G_V$, whereas the energy at $n=0$ stays
unaffected (see Ref.~\cite{Fukushima:2012mz} for a recent general
discussion of this point).

The effect of vector interactions on inhomogeneous phases has been 
investigated in Ref.~\cite{Carignano:2012sx}. In that analysis
the approximation was made to replace the density $n(z)$ in the 
mean-field Lagrangian by the spatial average $\nave$. 
This is a good approximation close to the 
restored phase and in particular at the Lifshitz point. The shape of the 
mass function at a given density is then independent of $G_V$, and the 
known analytical solutions for $G_V=0$ could basically be taken over. 
If we could apply the same approximation to our present analysis, 
we would obtain 
\beq
    \left.\frac{E}{N}\right|_{G_V} \;\approx\; 
    \left.\frac{E}{N}\right|_{G_V=0} \,+\ G_V\nave
    \;\equiv\; 
    \left(\frac{E}{N}\right)_< \,,
\label{eq:ENGVll}
\eeq
similar to the homogeneous case. 
This would further stabilize the minimum at $\nave =0$. 

It is obvious, however, that the replacement of $n$ by $\nave$ is not
a good approximation at low average densities where the quarks are 
strongly localized in the solitons and therefore feel a much stronger 
repulsion than suggested by $G_V\nave$. 
For instance, the energy of a single soliton is still enhanced by the
vector repulsion, even when the homogeneous background density and,
thus, the average density of the system goes to zero. 
Thus, the correction to $E/N$ should rather be obtained by 
integrating the local correction  to the energy density, 
$\delta\epsilon(z) = G_V n^2(z)$ over the volume and divide it by 
the integrated quark number density.
Since the integrals over the transverse area cancel, one obtains
\beq
    \left.\frac{E}{N}\right|_{G_V} 
    \;\approx\;
    \left.\frac{E}{N}\right|_{G_V=0} \,+\,
    G_V \frac{\int dz\, n^2(z)}{\int dz\, n(z)}
    \;\equiv\; 
    \left(\frac{E}{N}\right)_> \,.
\label{eq:ENGVul}
\eeq
This is still an approximation,
at least as long the density profiles $n(z)$ for $G_V=0$ 
which were given in Sec.~\ref{sec:dens} are used.
We expect that, in a fully self-consistent treatment, the vector repulsion 
between the quarks leads to a broadening of the density distribution,
which lowers the energy. \eq{eq:ENGVul} with the unmodified density 
profiles should therefore be taken as an upper limit of $E/N$,
while \eq{eq:ENGVll} provides a lower limit. 

Making use of the periodicity of the soliton lattice, \eq{eq:ENGVul}
can be simplified to
\beq
    \left(\frac{E}{N}\right)_> 
    \;=\; 
    \left.\frac{E}{N}\right|_{G_V=0} \,+\,
    \frac{G_V}{a\nave} \int\limits_0^a dz\, n^2(z)\,.
\label{eq:ENGV2}
\eeq
where $a = L/2$ is the distance between the solitons, introduced in 
\eq{eq:distance}.
For the single-soliton limit with $\muci < M_\mathit{vac}$,
we have $E/N|_{G_V=0} = \muci$, while
the integrals in \eq{eq:ENGVul} have readily been worked out in 
Sec.~\ref{sec:pss}.
This yields
\beq
    \left(\frac{E}{N}\right)_>\!\!\!(\nave = 0) \;=\;
    \muci \,+\,   
    G_V \ntyp\,, 
\label{eq:ENul0}
\eeq
with $\ntyp$ given in \eq{eq:ntyp}.

\begin{figure}
\begin{center}
\includegraphics[width=.45\textwidth]{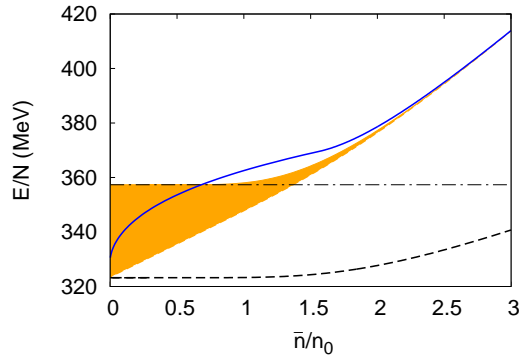}
\end{center}
\caption{Energy per particle as a function of the average density
for a vector coupling $G_V = G/2$.
The shaded area marks the range between the upper limit \eq{eq:ENGV2}
and the lower limit \eq{eq:ENGVll}.
The $\nave = 0$-value of the upper limit, \eq{eq:ENul0}, is denoted by the
dash-dotted line. 
Also shown are $E/N$ for homogeneous matter (solid line) and for 
inhomogeneous matter at $G_V=0$ (dashed line).
} 
\label{fig:EoN_Gv}
\end{figure}

In Fig.~\ref{fig:EoN_Gv} our results for $G_V = G/2$ are displayed as
functions of the average density. 
The range between the upper and lower limits of $E/N$ is indicated by
the shaded area. For comparison we also
show the results for homogeneous matter and for inhomogeneous matter
at $G_V=0$.
One can see that at high densities $(E/N)_>$,
$(E/N)_<$, and $E/N$ for homogeneous matter 
become practically degenerate.
This is not surprising, since in this regime the
amplitude $\Delta$ of the mass function becomes small and the 
density profile gets more and more washed out~\cite{Carignano:2010ac}.
At intermediate densities we find the energy of homogeneous matter 
to be higher than the upper limit of inhomogeneous matter, i.e.,
the inhomogeneous solution should be favored in this region. 

The situation is less clear at lower densities.
In the zero-density limit, $(E/N)_<$ and $(E/N)_\mathit{hom}$ 
converge against the corresponding limits without vector interactions, 
i.e., $\muci$ and $M_\mathit{vac}$, respectively, 
while $(E/N)_>$ approaches the value given in \eq{eq:ENul0}.
If the vector coupling is sufficiently small, 
\beq
    G_V < \frac{M_\mathit{vac}-\muci}{\ntyp} 
    = \frac{12\pi}{N_f N_c} \frac{1}{\muci}
    \left(\frac{1}{\muci} - \frac{1}{M_\mathit{vac}}\right)\,,
\eeq
the energy of homogeneous matter remains above the upper limit
for inhomogeneous matter, even at $\nave = 0$,
i.e., we can be rather sure that the inhomogeneous solutions stay favored.  
In the present example, however, $G_V$ is not so small, 
and we find the ordering 
$(E/N)_< < (E/N)_\mathit{hom} < (E/N)_>$
at low densities.
If the correct inhomogeneous solution is close to the upper limit,
this could mean that the ground state at low densities is homogeneous.
On the other hand, it is also possible that the inhomogeneous solution
remains favored if the solitons change their size in reaction to the
repulsive vector interaction.

\section{Conclusions}

In this article, we have studied the existence and the properties of 
self-bound quark matter in the NJL model at zero temperature,
focusing on inhomogeneous structures with one-dimensional spatial modulations.
The analysis was done in mean-field approximation.

For homogeneous matter, it was found long time ago that the model seems 
to allow for stable ``droplets'' of quark matter in the chirally restored 
phase if the interaction is sufficiently attractive. 
These droplets have vanishing pressure and a chemical potential lower 
than the vacuum constituent quark mass, so that they are in mechanical and 
chemical equilibrium with the vacuum. 
Related to this, they correspond to a minimum of the energy per particle 
as a function of density, so that they are stable against homogeneous
expansion or collapse. 
Neglecting finite size effects, this suggests to interpret these solutions
as quark bags, and the natural expectation would be that they have a spherical 
shape if surface effects are taken into account.

Allowing for one-dimensional inhomogeneities, however, it turns out that
the homogeneous droplets are unstable against forming a soliton lattice.
The solitons repel each other, so that the state with the lowest energy
per particle is reached at infinite lattice spacing, corresponding to a 
vanishing spatially averaged density. Inside the solitons, on the other
hand, the density is finite, roughly of the same order as in the homogeneous 
droplets. Their longitudinal size is about 1~fm,
determined by the inverse of the vacuum constituent quark mass.
Being one-dimensional objects embedded in the three-dimensional space,
the solitons are infinite in the two transverse directions. 
Thus, taking these results as they are, quark matter at low average
density should have a lasagne-like structure, with parallel plates
of high densities and voids in between. 

At this point, we should ask ourselves how these results can be
interpreted. In QCD, we expect that compressed quark matter, when it is
released, will expand and finally hadronize. 
At zero temperature, this means that the matter should split up into
baryons, each consisting of $N_c$ valence quarks.
These baryons may further interact with each other, forming nuclei or
nuclear matter, but keep their individuality as separate color-singlet 
objects.

In the NJL model, the ``droplet'' solutions found in the analysis of 
homogeneous quark matter have been suggested to be interpreted as schematic
baryons, since they are stabilized by the bag pressure and have a reasonable 
density.
Of course, strictly speaking, these solutions are infinite objects,
and a separation into finite baryons would require a negative surface 
tension~\cite{Alford:1997zt}, while recent analyses suggest that it is
positive~\cite{Molodtsov:2011ii,Lugones:2011xv,Pinto:2012aq}.
From this perspective, the one-dimensional solitons look like a step
in the right direction, as they are at least finite in one dimension, 
where they have a reasonable size. 
In particular, one might hope that the consideration of higher-dimensional 
inhomogeneous phases could reveal further instabilities, 
eventually leading to finite localized baryons as the true ground state
of matter at low densities. 

Unfortunately, this does not seem to be the case. 
Phases with two-dimensional modulations have been studied in 
Ref.~\cite{Carignano:2012sx} and were found to be disfavored against
one-dimensional modulations at low densities. Although the analysis was
restricted to sinusoidal shapes and certain parameters, 
it is unlikely that this will change if other shapes or parameters, 
or even three-dimensional modulations are considered. 
Nevertheless, more systematic studies in this direction are highly
desirable, in particular since at nonzero temperature one-dimensional 
periodic structures are known to be unstable against 
fluctuations~\cite{Landau:1969,Baym:1982}.
One should also revisit the old works
on the chiral quark soliton model~\cite{Alkofer:1994ph,Christov:1995vm,ripka}
and work out their relation to the present model.

Of course, there is a priori no reason to expect finite baryons 
to be the most favored objects in a nonconfining model.
On the other hand, the model predictions may still have some relevance
in the deconfined phase. 
The emergence of one-dimensional modulations can be understood
as a relic of the Peierls instability in $1+1$ dimensions~\cite{peierls},
which is a rather general mechanism.
Also the fact that the longitudinal size and the internal density of the
one-dimensional solitons are of the order to be expected for baryons
might indicate that confinement effects are not very drastic.
It is thus conceivable that lasagne-like patterns are preformed 
in expanding quark matter before hadronization takes place,
and it would be interesting to work out possible observable signatures. 

The present calculations could also be improved in several aspects:
In Sec.~\ref{sec:vec}, we gave only a lower and an upper limit for the
effect of vector interactions on the energy per particle. For the
upper limit, which is probably closer to the true solution,
we assumed that the density profiles remain unchanged when the vector 
interactions are switched on. 
However, we expect that the repulsive interaction leads to a broadening of 
the density peaks, which would lower the energy of the system.
In this way the solitons may continuously go over into homogeneous matter,
when the vector coupling is increased. 
We have also neglected the effect of spacelike vector condensates,
which should be present in anisotropic systems. 

Moreover, we should allow for BCS pairing of the quarks in the solitons. 
Inside the solitons we find densities of two to three times 
nuclear-matter density, for which gaps of the order of 50 to 100~MeV
have been found in homogeneous quark matter. It would be interesting to
see how this is changed for an inhomogeneous environment. 

If we want to extend our studies to quark matter in compact stars,
we must enforce beta equilibrium and electric neutrality.
This would put stress on the present solutions, since the chemical 
potentials and, hence, the favored periodicities of the soliton lattice 
would no longer be identical for up and down quarks.
If this effect is large, the system may find ways to accommodate
different periods, e.g., by forming a two-dimensional lattice, where the 
up- and down-quark condensates vary independently in different directions. 
It would also be interesting to include strange quarks and revisit the
problem of strange quark matter and strangelets in the 
NJL model~\cite{Buballa:1998pr}.

Unfortunately, these improvements of the model can no longer be done by
making use of the analytically known solutions of the $1+1$ dimensional
Gross-Neveu model, so that brute-force numerical diagonalizations of
the Hamiltonian seem to be unavoidable. 

Finally, we should also include fluctuations. 
An interesting scenario would be that they leave the inhomogeneous phase
(potentially with a higher-dimensional structure)
intact but turn the second-order phase transition from the vacuum phase
into first order. The minimum of $E/N$ would then be shifted to nonvanishing
average density. The resulting crystal could be a first step towards 
nuclear matter.

\section*{Acknowledgments}

We thank K.~Fukushima, E.-M.~Ilgenfritz, L.~von Smekal, M.~Thies, 
and J.~Wambach, for interesting discussions and valuable comments. 
We also thank the referee, whose questions helped us to identify
a mistake in Sect.~\ref{sec:sasi} of the original manuscript.
This work was partially supported by the Helmholtz Alliance EMMI, 
the Helmholtz International Center for FAIR, 
and by the Helmholtz Research School for Quark Matter Studies H-QM.


\end{document}